\documentstyle[preprint,aps]{revtex}
\begin{document}
\title{Quantum Phase Interference  for Quantum Tunneling in Spin Systems}
\author{J.--Q. Liang$^{1,2}$\footnote{e-mail:jqliang@mail.sxu.edu.cn},
 H. J. W. M\"{u}ller--Kirsten$^{1}
$\footnote{e-mail: mueller1@pysik.uni-kl.de},
 D. K. Park$^{1,3}$\footnote{e-mail:dkpark@genphys.kyungnam.ac.kr}
 and F.--C. Pu$^{4}$}
\address{1. Department of Physics, University of Kaiserslautern, D-67653 Kaiserslautern, Germany\\
2. Institute of Theoretical Physics and Department of Physics, Shanxi University, Taiyuan, Shanxi 030006, People's Republic of China\\
3. Department of Physics, Kyungnam University, Masan, 631-701, Korea\\
4. Institute of Physics and Center for Condensed Matter Physics, Chinese Academy of Sciences, Beijing 100080, People's Republic of China and Department of Physics,
Guangzhou Normal College, Guangzhou 510400, People's Republic of China}
\maketitle

\begin{abstract}
The point--particle--like Hamiltonian of a biaxial spin
 particle with external magnetic field along the hard axis is obtained in terms of the potential field description of
spin systems with exact spin--coordinate correspondence. The Zeeman energy term turns out to be an effective gauge potential which  leads to a nonintegrable phase of 
the Euclidean Feynman propagator.
 The phase interference
between clockwise and anticlockwise
 under barrier propagations  is recognized explicitly as the Aharonov-Bohm
effect. An additional phase which is significant for quantum phase interference is discovered with the quantum theory of spin systems besides the known
 phase obtained with the semiclassical treatment of spin. We also show the energy dependence of the effect and obtain the tunneling splitting at excited 
states with the help of periodic instantons.
\end{abstract}
\begin{center}
 PACS numbers 75.10.Jm, 75.30.Gw, 03.65.Sq, 75.45.+j, 11.10.Ef
\end{center}
\section{Introduction}
\label{sec:1}
Quantum tunneling in spin systems
 has attracted considerable attention
 both theoretically and experimentally
 in view of a possible experimental test of the tunneling effect
for mesoscopic single domain particles in
 which case it is known
 as macroscopic quantum
 tunneling\cite{leggett,gunther}. In
 particular the coherent tunneling between two 
degenerate metastable
orientations of magnetization results 
in the superposition of macroscopically
 distinguishable (classically degenerate) states,
the understanding of which is a longstanding problem in quantum
mechanics and is called macroscopic quantum
 coherence (MQC)\cite{garg1}. Till now only magnetic molecular
 clusters have been the most promising candidates to observe 
MQC\cite{wernsdorfer}.\\
Quenching of MQC for half-integer
 spin is a beautiful observation of quantum tunneling
 in spin particles  and has been investigated  in the literature
 by means of the phase interference of
spin coherent state-Feynman-paths which possess a phase with obvious geometric meaning\cite{gunther,loss,chudnovsky1,prokofev,liang1}. The quenching of MQC has been 
interpreted physically as Kramers' degeneracy. The
geometric phase
has  also been shown to be equivalent to a Wess--Zumino type
 interaction in quantum mechanics\cite{liang2}. However the effect of geometric phase
interference is far richer than  Kramers' degeneracy. For example, the Zeeman energy of an external magnetic field applied along the hard axis for a biaxial spin particle can be 
introduced to produce an additional geometric phase\cite{garg2}. The resulting quenching of the tunneling splitting or MQC is in this case not related to Kramers' degeneracy since the
external magnetic field breaks the time reversal
 symmetry. A more detailed investigation of quantum phase interference has been given recently\cite{kou}. An experimental observation of
the  magnetic field dependent oscillation of tunneling 
splitting has also been reported\cite{wernsdorfer} 
for the octanuclear iron oxo-hydroxo
 cluster $Fe_8$. The giant spin model we consider
here is suitable to describe
 the $Fe_8$ molecular cluster and therefore has practical interest.\\
In the traditional theory, the quantum phase induced by the Zeeman term has been investigated with the semiclassical method in which the spin is treated as a classical vector. With the help of
the spin coherent state-path-integral
 technique an effective
 Hamiltonian and Lagrangian are
 obtained. The Zeeman term is  proportional
 to the linear velocity  and therefore emerges as a
phase of the Feynman kernel in the imaginary time for quantum tunneling. 
It is, no doubt, interesting
 to explore the  underlying physics of
 the  phase related to the Zeeman energy of the
 magnetic field and  to present an analysis
 of the quantum phase based
on a full quantum mechanical theory of the spin system.
 To this end we  use a recently developed
 method, namely, the potential
 field description of quantum spin systems
 of Ulyanov and Zaslavskii (UZ)\cite{ulyanov} and begin with the
Schr\"{o}dinger equation of the spin
 particle. In the UZ method the spin-coordinate correspondence is exact
 unlike the semiclassical approach
 of spin where the correspondence is approximate
in the large spin limit
 (see Appendix 1 for details). A point-particle-like
 Hamiltonian is obtained
 and the Zeeman term of the
 magnetic field along the hard axis of the biaxial spin system becomes
a gauge potential which does not affect the equation of motion. However, the nonintegrable phase of the gauge potential leads to  
 the quantum phase interference
 known as the Aharonov-Bohm(AB) effect. A substantial
 result derived from the quantum theory
 of spin is the additional phase induced by the Zeeman term
 which is
significant to the quantum phase interference
 and is overlooked in the semiclassical approach. The tunneling splittings for both ground state and excited states are also obtained up to the one loop
approximation.
The paper is organized as follows.
 In Sec. 2  we first give a brief  review of the
 semiclassical treatment of
  spin in quantum tunneling. The effective Hamiltonian like that of a point particle  
for a biaxial spin particle is obtained by
 starting from the Schr\"{o}dinger equation
following the potential field description of  quantum spin systems. We show how the Zeeman energy term of the magnetic filed along hard axis becomes a gauge potential. The tunneling
splitting and its oscillation with respect to
 the magnetic field are discussed in Sec. 3.
  We obtain in Sec. 4 the periodic
 instantons and oscillation of tunneling splitting at
excited states and demonstrate  the energy
 dependence of the oscillation. Our conclusions
 and discussions are given in Sec. 5.\\

\section{Effective potentials of the biaxial spin particle in a magnetic field}
\label{sec:2}
We consider a giant spin model which is assumed to have biaxial anisotropy with XOY easy plane and the easy y-axis in the XOY-plane. An external magnetic field is applied
along the hard z-axis. The Hamilton operator of the model can be written as
\begin{equation}
\label{1}
\hat{H}=K_{1}\hat{S}^{2}_{z}+K_{2}\hat{S}^{2}_{x}-g\mu_{B}h\hat{S}_{z}
\end{equation}
where $\hat{S}_{x}$, $\hat{S}_y$ and $\hat{S}_{z}$ are the three components of the spin operator.  $K_{1}$ and $K_{2}$ with $K_{1}>K_{2}>0$ are the anisotropy constants and
$\mu_{B}$ is the Bohr magneton.
 $g$ is the spin $g$-factor which is taken to be 2 here. The
 last term of the Hamiltonian is the Zeeman energy
 associated with the magnetic
field h. The Hamiltonian eq.(1) is believed
 to describe the $Fe_{8}$ molecular cluster and is the same as that in ref.[4]. Before we begin with the investigation with the UZ method we  give a brief
review of the semiclassical approach for the model of eq.(1). In the semiclassical method the spin is treated as a classical vector
\begin{equation}
\label{2}
{\bf S}=s(\sin\theta\cos\phi, \sin\theta\sin\phi, \cos\theta)
\end{equation}
The spin-coordinate correspondence is given by  the definition of canonical variables\cite{perelomov} $\phi$ and $p=s\cos\theta$. As shown in Appendix 1 the usual spin commutation
relation $[\hat{S}_{i}, \hat{S}_{j}]=i\epsilon_{ijk}\hat{S}_{k}$
 is only approximately recovered in the
 large spin s limit. With the spin coherent state path integral
technique one obtains an effective Hamiltonian\cite{liang1,kou,liang3}
\begin{equation}
\label{3}
H_{s}=\frac{p^{2}}{2m(\phi)}-\alpha p+V(\phi)
\end{equation}
with
\begin{equation}
\label{4}
m(\phi)=\frac{1}{2K_{1}(1-\lambda\sin^{2}\phi)},\hspace{0.5cm}V(\phi)=K_{2}s^{2}\sin^{2}\phi,\hspace{0.5cm}\lambda=\frac{K_{2}}{K_{1}},\hspace{0.5cm}\alpha=g\mu_{B}h
\end{equation}
and Lagrangian
\begin{equation}
\label{5}
L_{s}=\frac{m(\phi)}{2}(\dot{\phi}+\alpha)^{2}-V(\phi)
\end{equation}
The position dependent mass may
 create an ordering ambiguity upon
 quantization. This is the reason why we
 use the elliptic integral transformation 
in the following to obtain a point--particle--like 
Hamiltonian with a constant mass. In the above
 derivation we have shifted the angle $\phi$ by $\frac{\pi}{2}$ for our convenience. The periodic potential $V(\phi)$ has degenerate
vacua. The quantum tunneling from one vacuum ($\phi=0$) to the
 neighboring one ($\phi=\pi$) is dominated by  
instantons and evaluated to exponential accuracy by
\begin{equation}
\label{6}
e^{-S_{sc}}=e^{-\int L^{e}_{s}d\tau}
\end{equation}
where
\begin{equation}
\label{7}
L^{e}_{s}=\frac{m(\phi)}{2}(\frac{d\phi}{d\tau})^{2}-i\alpha m(\phi)\frac{d\phi}{d\tau}+V(\phi)-\frac{m(\phi)}{2}\alpha^{2}
\end{equation}
is the Euclidean Lagrangian with
 the imaginary time $\tau=it$. The imaginary
 part (the second term) in $L^{e}_{s}$ induced by 
the Zeeman term becomes a phase in eq.(6)
\begin{equation}
\label{8}
e^{-S_{sc}}=e^{-\tilde{S}_{sc}}e^{i\theta_{s}}
\end{equation}
where $\tilde{S}_{sc}$ is the remaining
 action, and the phase derived with the semiclassical method is seen to be
\begin{equation}
\label{9}
\theta_{s}=\int_{0}^{\pi}\alpha m(\phi)d\phi
=\frac{\alpha\pi}{2K_{1}\sqrt{1-\lambda}}
\end{equation}
which leads to the quantum phase interference between
 clockwise and anticlockwise tunnelings. Since we
 here emphasize the phase induced by
 the Zeeman energy term, the known phase term
$s\frac{d\phi}{d\tau}$ (responsible for the spin
 parity effect) which we omitted in the Euclidean action should be understood. We now turn to the quantum theory of spin.\\
   
Following ref.[12] we start from the Schr\"{o}dinger equation
\begin{equation}
\label{10}
\hat{H}\Phi(\phi)=E\Phi(\phi)
\end{equation}
The explicit form of the action of the spin operator on the function $\Phi(\phi)$ is seen to be
\begin{equation}
\label{11}
\hat{S}_{x}=s\cos\phi-\sin\phi\frac{d}{d\phi},\hspace{0.5cm}  \hat{S}_{y}=s\sin\phi+\cos\phi\frac{d}{d\phi},\hspace{0.5cm}  \hat{S}_{z}=-i\frac{d}{d\phi}
\end{equation}
where the generating function $\Phi(\phi)$ is constructed in terms of the conventional spin functions of the $\hat{S}_{z}$  representation such as
\begin{equation}
\label{12}
\Phi(\phi)=\sum^{s}_{m=-s}\frac{c_{m}}{\sqrt{(s-m)!(s+m)!}}e^{im\phi}
\end{equation}
which obviously obeys the following boundary condition
\begin{equation}
\label{13}
\Phi(\phi+2\pi)=e^{2\pi is}\Phi(\phi)
\end{equation}
Thus we have periodic wave functions for integer spin s and antiperiodic functions for half-integer s. The antiperiodic wave functions naturally give rise to the spin parity
effect  as we shall see. Substitution of the differential spin operators eq.(11) into eq.(10) yields
\begin{equation}
\label{14}
[-K_{1}(1-\lambda\sin^{2}\phi)\frac{d^{2}}{d\phi^{2}}-K_{2}(s-\frac{1}{2})\sin2\phi\frac{d}{d\phi}+i\alpha\frac{d}{d\phi}+V(\phi)]\Phi(\phi)=E\Phi(\phi)
\end{equation}
with
\begin{equation}
\label{15}
V(\phi)=K_{2}s^{2}\cos^{2}\phi+K_{2}s\sin^{2}\phi
\end{equation}
In the new variable x defined by
\begin{equation}
\label{16}
x=\int^{\phi}_{0}\frac{d\phi^{\prime}}{\sqrt{1-\lambda\sin^{2}\phi}}=F(\phi,k)
\end{equation}
which is the incomplete elliptic integral of the first kind
 with  modulus $k^{2}=\lambda$, the trigonometric functions $\sin\phi$ and $\cos\phi$ become the Jacobian elliptic
functions sn(x), cn(x) with the same modulus respectively. We then make use of the following transformation,
\begin{equation}
\label{17}
\Phi(\phi(x))=dn^{s}(x)e^{if(x)}\psi(x)
\end{equation}
where $dn(x)=\sqrt{1-\lambda sn^{2}(x)}$ is also a Jacobian elliptic function.
Substituting eq.(17) into eq.(14) we obtain, after some tedious but not too complicated algebra, an equivalent Schr\"{o}dinger equation with the desired Hamiltonian, i.e.
\begin{equation}
\label{18}
\{K_{1}[-i\frac{d}{dx}+A(x)]^{2}+U(x)\}\psi(x)=E\psi(x)
\end{equation}
The function f(x) in the unitary transformation
 is determined by the requirement of gauge covariance and the scalar
potential is required to be real. f(x) is therefore defined by
\begin{equation}
\label{19}
\frac{df(x)}{dx}=\frac{-\alpha s}{K_{1}dn(x)}
\end{equation}
A gauge potential induced by the Zeeman energy term is found to be
\begin{equation}
\label{20}
A(x)=-\frac{\alpha(2s+1)}{2K_{1}dn(x)}
\end{equation}
 The scalar potential
\begin{equation}
\label{21}
U(x)=\xi cd^{2}(x), \hspace{1cm} cd(x)=\frac{cn(x)}{dn(x)}
\end{equation}
is periodic with period $2{\cal K}(k)$ and symmetric with respect to the coordinate origin $x=o (U(x)=U(-x))$. The quantity ${\cal K}(k)$ is the complete elliptic integral of the first kind. 
The minima of the potential, namely, the vacua which have been shifted to zero by adding a constant, are located at $\pm (2n+1){\cal K}(k)$ with n being an integer. The positions of potential
 peaks are at $\pm 2n{\cal K}(k)$, and barrier height is
\begin{equation}
\label{22}
\xi=K_{2}s(s+1)+\frac{\lambda \alpha^{2}}{4(1-\lambda)K_{1}}
\end{equation}
When $\alpha=0$, i.e. the Zeeman term in eq.(1) vanishes, the potential becomes exactly the same as that in ref.[15]. The shape of the scalar potential  is not changed by the external
magnetic field along the hard axis. In the new variable x the wave function $\Phi (\phi (x))$ is also periodic for integer s and antiperiodic for half-integer s with a period $4{\cal K}(k)$ and the boundary 
condition of the wave function
 $\psi (x)$ is, however, determined by eqs.(17) and (19), i.e.
\begin{equation}
\label{23}
\psi(x+4{\cal K}(k))=(-1)^{2s}e^{i\frac {2\alpha s\pi}{K_{1}\sqrt{1-\lambda }}}\psi(x)
\end{equation}
One should bear in mind from eq.(16) that $x={\cal K}(k)$ corresponds to the original angle variable $\phi=\frac{\pi}{2}$. The boundary condition eq.(23) plays an important role in the following calculation of the 
tunneling splitting.\\

\section{Tunneling splitting at the ground state}
\label{sec:3}
 The tunneling between degenerate vacua
 (the case we consider here) results in the level
 splitting and is dominated by (vacuum) instantons which 
are nontrivial solutions of the Euclidean
 equation of motion with finite action. In the
 context of quantum mechanics the instanton
 may be visualized as a pseudoparticle moving between
degenerate vacua under the barrier and has
 nonzero topological charge but zero energy.
 The tunneling splitting can be obtained from
 the transition amplitude between degenerate vacua which has a
Euclidean path--integral representation.
 The first explicit calculation of
 the tunneling splitting  in terms of the
instanton method was carried
 out long ago for the double-well potential\cite{gildener}.\\
The Hamilton function corresponding to eq.(18) is
\begin{equation}
\label{24}
H=\frac{1}{2m}[P+A(x)]^{2}+U(x)
\end{equation}
where $m=\frac{1}{2K_{1}}$ is the mass of the point-like particle. The Lagrangian is
\begin{equation}
\label{25}
L=\frac{m}{2}\dot{x}^{2}-A(x)\dot{x}-U(x)
\end{equation}
With the Wick rotation $\tau=it$ the Euclidean Lagrangian is seen to be
\begin{equation}
\label{26}
L_{e}=\frac{m}{2}\dot{x}^{2}+iA(x)\dot{x}+U(x)
\end{equation}
In the above Euclidean
 Lagrangian and from
 now on $\dot{x}=\frac{dx}{d\tau}$ denotes
 the imaginary time derivative. The gauge potential $A(x)$
 indeed does not affect the equation of motion which is 
\begin{equation}
\label{27}
m\ddot{x}=\frac{dU(x)}{dx}
\end{equation}
However, it leads to the phase interference  which can be observed as the oscillation of tunneling splitting. This is exactly an AB effect in a generalized
 meaning.\\
The instanton solution of eq.(27) can be found by direct integration and the result is
\begin{equation}
\label{28}
x_{c}(\tau)=sn^{-1}(\tanh\omega\tau),\hspace{1cm}\omega^{2}=4K_{1}\xi
\end{equation}
which is nothing but a kink configuration. The instanton starts  from the vacuum  $x_{i}=-{\cal K}(k)$ at $\tau=-\infty$ and reaches the centre of the potential barrier ($x=0$)
at $\tau =0$ and then arrives at the neighboring vacuum $x_{f}={\cal K}(k)$ at $\tau =\infty$. The Euclidean action evaluated along the instanton trajectory is
\begin{equation}
\label{29}
S_{c}=\int_{-\infty}^{\infty}L_{e}(x_{c}(\tau), \dot{x}_{c}(\tau))d\tau =B-i(2s+1)\theta_{s}
\end{equation}
where the first term
\begin{equation}
\label{30}
B=\sqrt{\frac{\xi}{K_{2}}}\ln\frac{1+\sqrt{\lambda}}{1-\sqrt{\lambda}}
\end{equation}
 reduces to the well known action\cite{liang1,kou,liang3,enz1,enz2,chudnovsky2} when $\alpha=0$. The imaginary part leads to a phase in the Euclidean Feynman propagator which is
$2s+1$ times the semiclassical phase $\theta_{s}$.\\

To calculate the tunneling splitting,
 we start from the instanton induced transition amplitude,
\begin{equation}
\label{31}
\langle x_{f}(\beta)\vert x_{i}(-\beta)\rangle =\sum_{m_{f}, n_{i}}\langle x_{f}\vert m_{f}\rangle\langle m_{f}\vert \hat{P}_{E}e^{-\beta \hat{H}}\vert n_{i}\rangle\langle
n_{i}\vert x_{i}\rangle
\end{equation}
$\hat {P}_{E}$ is the
 projector onto the subspace
 of fixed energy\cite{kuznetsov}
 and $\vert n_{i}\rangle , \vert m_{f}\rangle $ are the excitations above two vacua lying on different 
sides of the
barrier. The left hand side
 of eq.(31) has the path integral
 representation and is evaluated
 in the following. We consider
 the tunneling from initial vacuum $x_{i}=-{\cal K}(k)$
(corresponing to the original
 angle variable $\phi_{i}=-\frac{\pi}{2}$) to
 the neighboring one $x_{f}={\cal K}(k)$ 
($\phi_{f}=\frac{\pi}{2}$) for the fixed energy $E_{0}$ which is
the degenerate ground
 state energy. The small tunneling
 splitting of the ground state is obtained from eq.(31) such that
\begin{equation}
\label{32}
\Delta E_{0}\sim \vert \frac{e^{2\beta E_{0}}}{\beta}F\int_{-{\cal K}(k)}^{{\cal K}(k)}{\cal D}{x}e^{-\int_{-\beta}^{\beta}L_{e}d\tau}\vert
\end{equation}
where
\begin{equation}
\label{33}
F=\frac{1}{\psi_{0}({\cal K}(k))\psi_{0}^{*}(-{\cal K}(k))}=\frac{e^{-i(\pi s+2s\theta_{s})}}{N},\hspace{1cm}    N=\psi_{0}(0_{f})\psi_{0}(0_{i})
\end{equation}
The second equality in F comes from the boundary condition of our wave function eq.(23) and $0_{i}, 0_{f}$ denote the coordinate origins of the local frames associated with each
potential well. N is
 then a normalization constant
 calculated with the harmonic oscillator
 approximated wave function
 of the ground state. Substitution of the Lagrangian eq.(26)
into the Feynman kernel in eq.(32) yields
 our interesting phase,
\begin{equation}
\label{34}
\int{\cal D}{x}e^{-\int_{-\beta}^{\beta}L_{e}d\tau}=e^{i(2s+1)\theta_{s}}G(x_{f},\beta ; x_{i}, -\beta )
\end{equation}
It is somewhat surprising
 that the quantum theory gives rise to $2s+1$ times the
 semiclassical phase angle $\theta_{s}$ instead of just one. In Appendix 2 we explain the reason why the significant
phase angle $2s\theta_{s}$ 
can be missed in the semiclassical
 treatment of spin based on
 the large spin s
 limit. However, the additional phase, i. e. $2s\theta_{s}$, in
 the Euclidean Feynman kernel is cancelled by the
the phase of F in eq.(33) and does not affect the
 tunneling splitting of the ground
 state. We will see in the following
 section that the cancellation would not
 be exact for excited states and
 the additional phase has effect on the tunneling splitting\\
We should bear in  mind that the above
 phase is obtained  by an anticlockwise tunneling.
 The remaining Feynman kernel
\begin{equation}
\label{35}
G=\int{\cal D}{x}e^{-\int_{-\beta}^{\beta}L^{\prime}_{e}d\tau},\hspace{1cm}  L^{\prime}_{e}=\frac{m}{2}\dot{x}^{2}+U(x)
\end{equation}
is independent of the direction
 of tunneling. For the clockwise tunneling from
 the same initial position $-{\cal K}(k)$ to the
 final position $-3{\cal K}(k)$  the result is the same except with
an opposite sign  of the phase. Adding 
 the  contributions of clockwise
 and anticlockwise tunnelings
 we finally have the tunneling splitting expressed as
\begin{equation}
\label{36}
\Delta E_{0}\sim\frac{e^{2\beta E_{0}}}{\beta N}\vert \cos[\pi s+\theta_{s}]\vert G(x_{f}, \beta; x_{i}, -\beta)
\end{equation}
The tunneling kernel G in the one loop approximation,
 namely including the preexponential factor,
 can be calculated with the standard procedure. Before
 we give the final result, it is
worthwhile to point out that in the evaluation
 of G the contributions from one instanton
 and one  instanton plus the infinite
 number of instanton-anti-instanton
 pairs will be taken into
account. However, the phase induced by the gauge
 potential for an instanton-anti-instanton pair
 vanishes. Thus the single
 instanton phase is factored out off the tunneling kernel. We have
\begin{equation}
\label{37}
G\sim 2N\beta e^{-2\beta E_{0}}Qe^{-B},\hspace{1cm}  Q=2^{\frac{5}{2}}[\frac{\xi^{\frac{3}{2}}K_{1}^{\frac{1}{2}}}{(1-\lambda)\pi}]^{\frac{1}{2}}
\end{equation}
where $N=\frac{1}{\sqrt{2\pi}}
(\frac{\xi}{K_{1}})^{\frac{1}{4}}$,
 and $E_{0}=\frac{\omega}{2}$ is the usual ground state
 energy of the harmonic oscillator.
 The tunneling splitting is thus\cite{liang3}
\begin{equation}
\label{38}
\Delta E_{0}
=\vert\cos[\pi s+\theta_{s}]\vert 4\Delta\varepsilon_{0}, \hspace{1cm}
 \Delta\varepsilon_{0}=Qe^{-B}
\end{equation}
When the external magnetic field
 vanishes ($\alpha=0$) the tunneling
 splitting reduces exactly
 to the previous result\cite{liang1,liang3} except
  that $s^{2}$ in the splitting amplitude of the semiclassical 
treatment is corrected
 as $s(s+1)$ by the quantum theory of spin. The well known spin parity effect 
(namely, the tuuneling splitting
would be quenched for half--integer spin s)
 is recovered by the factor
 $\vert\cos\pi s\vert$ and is
 surely due to the antiperiodicity
 of the wave function in the
 quantum theory of spin. 
 The tunneling splitting oscillates with
 the external field h and vanishes when
\begin{equation}
\label{39}
s\pi+\theta_{s}=
\pi[s+\frac{\mu_{B}gh}{2K_{1}\sqrt{1-\lambda}}]=(n+\frac{1}{2})\pi
\end{equation}
where $n$ is an integer.
 The oscillation period
 of the tunneling splitting
 with respect to the external field h  is given by
\begin{equation}
\label{40}
\Delta h=\frac{2K_{1}\sqrt{1-\lambda}}{g\mu_{B}}
\end{equation}
To verify the validity  of
 the tunneling splitting
 eq.(38) we compare
 the splitting value
 of eq.(38)  as a function
 of the external magnetic
 field h with the numerical
 result obtained by performing
 a diagonalization of the Hamilton operator
eq.(1). Adopting  the data of the anisotropy
 contants given in ref.[4]
 such that $D=0.292K$, $E=0.046K$ and taking into account
 the relations between anisotropy constants
 $K_{1}$, $K_{2}$ and D, E, i.e. $K_{1}=D+E$, $K_{2}=D-E$, the 
oscillating amplitude of the tunneling
 splitting calculated from eq.(38) which
 begins from $6.286\times 10^{-10}K$ for $s=10$ and
 increases with the magnetic
 field agrees with the numerical value
 of diagonalization perfectly.
 The period is 
$\Delta h=0.26T$ which is
 substantially smaller than the experimental
 value 0.4T\cite{wernsdorfer}. It
 has been pointed out that
 the discrepancies between
 experimental and theoretical results
 can be resolved by including
 higher order terms of $\hat S_{z}$
and $\hat S_{x}$ in the Hamilton operator eq.(1)
 besides the quadratic terms\cite{wernsdorfer}.\\

\section{Tunneling splitting and quantum phase interference at excited states}
\label{sec:4}
 The quantum phase induced by the Zeeman term is manifestly computed from the Euclidean Feynman paths between two turning points which depend on energy. We now investigate the tunneling
and related quantum phase interference at excited states. The starting point is again the  transition amplitute of the barrier penetration  projected onto the subspace of fixed energy, i.e.
\begin{equation}
\label{41}
\sum_{m,n}<E^{f}_{n}\vert\hat{P}_{E}e^{-2\beta\hat{H}}\vert E^{i}_{m}>=\int dx_{f}dx_{i}\psi^{*}_{E}(x_{f})\psi_{E}(x_{i})G(x_{f}, \beta; x_{i},-\beta)
\end{equation}
from which the tunneling splitting is written as,
\begin{equation}
\label{42}
\Delta E\sim \frac{e^{2E\beta}}{\beta}\vert e^{-is(\pi -2\theta_{s})}\int d\tilde{x}_{f}d\tilde{x}_{i}\psi_{E}(\tilde{x}_{f})\psi_{E}(\tilde{x}_{i})G\vert
\end{equation}
where
 $\tilde{x}_{i}={\cal K}(k)+x_{i}$,
 $\tilde{x}_{f}=-{\cal K}(k)+x_{f}$
 denote the coordinates in the local frames with origins at $-{\cal K}(k)$ and ${\cal K}(k)$ respectively. Thus
the phase factor of our wave function
 $\psi_{E}$ can be factorized out. The tunneling at finite energy
 E is dominated by the periodic 
instanton\cite{manton,liang4} which
 satisfies the following 
integrated equation of motion,
\begin{equation}
\label{43}
\frac{m}{2}\dot{x}^{2}-U(x)=-E
\end{equation}
with periodic boundary condition. The periodic instanton is found to be
\begin{equation}
\label{44}
x_{c}(\tau)=cd^{-1}\left(\sqrt{\frac{[\sin^{-1}sn(\tilde{\omega}\tau,\tilde{k})]^{2}(1-\eta^{2})-(1-\lambda\eta^{2})}{\lambda[\sin^{-1}sn(\tilde{\omega}\tau,
\tilde{k})]^{2}(1-\eta^{2})-(1-\lambda\eta^{2})}},\tilde{k}\right)
\end{equation}
where
\begin{equation}
\label{45}
\tilde{\omega}=2\sqrt{K_{1}\xi(1-\lambda\eta^{2})},\hspace{0.5cm} \eta=\sqrt{\frac{E}{\xi}},\hspace{0.5cm} \tilde{k}^{2}=\frac{1-\eta^{2}}{1-\lambda\eta^{2}}
\end{equation}       
The periodic instanton moves between
 two turning points $x_{\pm}$ depending on energy
\begin{equation}
\label{46}
x_{\pm}=\pm cd^{-1}(\eta, \tilde{k})
\end{equation}
When the energy tends to zero
 the periodic instanton reduces 
to the vacuum instanton of eq.(28). The
 Euclidean action evaluated along the
 periodic instanton is
\begin{equation}
\label{47}
S_{c}=W+2E\beta-i\theta_{E}
\end{equation}
where
\begin{equation}
\label{48}
W=\int_{-\beta}^{\beta}m\dot{x}^{2}_{c}d\tau=2\eta^{2}\sqrt{\frac{\xi}{K_{1}(1-\lambda\eta^{2})}}[\Pi({\eta^{\prime}}^{2}, \tilde{k})-{\cal K}(\tilde{k})]
\end{equation}
with ${\eta^{\prime}}^{2}=1-\eta^{2}$, where $\Pi$ denotes the complete elliptic integral of the third kind. 
The tunneling phase for the
 anticlockwise tunneling (from $x_{-}$ to $x_{+}$) is seen to be
\begin{equation}
\label{49}
\theta_{E}=\int_{x_{-}}^{x_{+}}A(x_{c})dx_{c}=\frac{(2s+1)\alpha}{K_{1}\sqrt{1-\lambda}}[\tan^{-1}\frac{\eta^{\prime}-\eta}{\eta^{\prime}+\eta}+\frac{\pi}{4}]
\end{equation}
which tends to the vacuum
 instanton phase when
 $E=0$ ($\eta=0$, $\eta^{\prime}=1$). The clockwise
 tunneling gives rise to the same
 result except for the phase with an
 opposite sign. Adding  the
two classes of the tunneling
 kernels the level splitting is seen to be
\begin{equation}
\label{50}
\Delta E\sim\frac{e^{2E\beta}}{\beta}|\cos(s\pi +\theta_{E}-2s\theta_{s})|I
\end{equation}
where
\begin{equation}
\label{51}
I=\int d\tilde{x}_{f}d\tilde{x}_{i}\psi_{E}(\tilde{x}_{f})\psi_{E}(\tilde{x}_{i})\tilde{G}
\end{equation}
The term $2s\theta_{s}$ in eq.(50) comes from
 the boundary condition
 of  $\psi $ eq.(23). The difference, i.e. $\theta_{E}-2s\theta_{s}$,
 is not just a simple semiclassical
 phase $\theta_{s}$ in this case. The phase independent
 tunneling kernel $\tilde{G}$ is now
  evaluated with the help of the periodic
 instanton. Following the procedure
 in refs.[23] and [24] we take  into account the contributions
 of the instanton and
instanton plus the infinite number of
 pairs and compute the end point integrations
 with the help of WKB wave functions for $\psi_{E}$.
 A quite general formula for eq.(51) is
\begin{equation}
\label{52}
I\sim 2\beta e^{-2E\beta}[\frac{1}{4{\cal K}(\hat{k})}]e^{-W}, \hspace{1cm} \hat{k}^{2}=\frac{(1-\lambda)\eta^{2}}{1-\lambda\eta^{2}}
\end{equation}
The level splitting is then given by
\begin{equation}
\label{53}
\Delta E=|\cos(s\pi+\theta_{E}-2s\theta_{s})
|\frac{1}{{\cal K}(\hat{k})}e^{-W}
\end{equation}
For low lying excited states ($\eta<<1$, $\hat{k}<<1$, $\tilde{k}^{\prime}=\sqrt{1-\tilde{k}^{2}}<<1)$ the energy E may be replaced by harmonic oscillator approximated eigenvalues 
$E\rightarrow E_{n}=
(n+\frac{1}{2})\omega $. Expanding
the complete elliptic integrals 
$\Pi({\eta^{\prime}}^{2}, \tilde{k})$,
 ${\cal K}(\tilde{k})$ in W (eq.(48)) as power series of $\tilde{k}^{\prime}$ and ${\cal K}(\hat{k})$ in eq.(52) as power 
series of $\hat{k}$ we obtain after
 some tedious algebra the tunneling splitting of the nth excited state, i.e.
\begin{equation}
\label{54}
\Delta E_{n}=|\cos(s\pi+\theta_{E_{n}}-2s\theta_{s})|4\Delta\varepsilon_{n}
\end{equation}
where
\begin{equation}
\label{55}
\Delta\varepsilon_{n}
=\frac{2^{3n}}{n!(1-\lambda)^{n}}
(\frac{\xi}{K_{1}})^{\frac{n}{2}}\Delta\varepsilon_{0}
\end{equation}
In eq.(54) $\theta_{E_{n}}$ denotes 
the phase angle at nth excited state which
 is obtained from eq.(49) with replacing
 the energy E by $(n+\frac{1}{2})\omega$. 
When $\alpha=0$ the tunelling
 splittings at excited states again
 coincide with the previous results\cite{liang1,zhang}
 in terms of the semiclassical
 treatment of spin in large spin limit 
which means that  the difference 
between $s^{2}$ and $s(s+1)$ can be neglected.

\section{conclusion}
\label{sec:5}  
On the basis of the UZ method for quantum
 spin systems we found that the Zeeman term of
 the external magnetic field along
 the hard axis for a biaxial spin particle
 indeed turns out to be a gauge potential
in the point-particle-like Hamilton
 operator. The gauge potential does not
 affect the equation of motion but leads to quantum
 phase interference as an AB type effect
 in the spin tunneling.
An additional phase angle $2s\theta_{s}$ of
 the Euclidean action obtained by
 means of the quantum mechanical
 treatment of spin does not affect
 the tunneling splitting of the ground state but
 has effect on the tunneling splitting of excited states.
In addition the splitting amplitude 
is modified by the quantum theory of spin. We present
 a formula of the tunneling
 splitting, eq.(54), as a function
 of the magnetic field, valid for low lying 
excited states, which
for molecular clusters in which the total
 spin is only about ten 
 is more 
accurate than  the semiclassical treatment of spin
for describing the quantum tunneling.\\
\vspace{1cm}

{\bf Acknowledgment:} This work was supported
 by the National Natural Science Foundation 
of China under Grant Nos. 1967701 and 19775033. 
J.-Q. L. and D. K. P. also acknowledge support by the
Deutsche Forschungsgemeinschaft.
\vspace{1cm}

\begin{center}
{\bf {Appendix 1:Approximate
 spin-coordinate correspondance in the semiclassical treatment of spin}}
\end{center}
In the conventional application of
 the spin coherent state technique, two canonical
 variables, $\phi$ and $p=s\cos\theta$ are adopted with the usual quantization
\begin{equation}
[\phi, p]=i \eqnum{A1}
\end{equation}
We show in the following that the
 spin-coordinate correspondence
 is only approximate up to order $0(s^{-3})$.\\
From the relation between the spin operators
 and the polar coordinate angles
\begin{equation}
S_{x}=s\sin\theta\cos\phi,\hspace{1cm}S_{y}
=s\sin\theta\sin\phi,\hspace{1cm}S_{z}=s\cos\theta \eqnum{A2}
\end{equation}
the usual commutation relation of spin operators reads
\begin{equation}
[S_{x}, S_{y}]=s^{2}
[\sin\theta\cos\phi, \sin\theta\sin\phi]
=s^{2}\sin\theta[\cos\phi,\sin\theta]\sin\phi
+s^{2}\sin\theta[\sin\theta,\sin\phi]\cos\phi \eqnum{A3}
\end{equation}
Using eq.(A1), one can prove the following relations
\begin{equation}
[\sin\theta,\cos\phi]
=A_{+}\cos\phi+iA_{-}\sin\phi,\hspace{1cm}[\sin\theta,\sin\phi]=A_{+}\sin\phi-iA_{-}\cos\phi \eqnum{A4}
\end{equation}
with
\begin{eqnarray*}
A_{+}=\frac{1}{2}\left(\sqrt{1-(\cos\theta+\gamma)^{2}}
+\sqrt{1-(\cos\theta-\gamma)^{2}}\right),\\
A_{-}=\frac{1}{2}\left(\sqrt{1-(\cos\theta+\gamma)^{2}}
-\sqrt{1-(\cos\theta-\gamma)^{2}}\right)
\end{eqnarray*}
where $\gamma=\frac{1}{s}$. 
Substituting eq.(A4) into eq.(A3), one has
\begin{equation}
[S_{x}, S_{y}]=-is^{2}\sin\theta A_{-}
=i\gamma s^{2}\cos\theta +0(\gamma^{3}) \eqnum{A5}
\end{equation}
i.e.
\begin{equation}
[S_{x},S_{y}]=iS_{z}+0(s^{-3}) \eqnum{A6}
\end{equation}
which implies that the usual
 commutation relation holds only in the large spin limit.
\vspace{1cm}

\begin{center}
{\bf {Appendix 2: Recovering  the semiclassical
 phase in the large s limit}}
\end{center}
To understand the reason why the
 phase angle $2s\theta_{s}$  is missed in the
 semiclassical treatment of spin we consider the
 following Schr\"odinger equation obtained without 
the unitary transformation $e^{if(x)}$ in the
 transformation eq.(17) for our spin sytem:
\begin{equation}
[-K_{1}\frac{d^{2}}{dx^{2}}
+i\frac{\alpha}{dn(x)}\frac{d}{dx}
-is\alpha\lambda\frac{sn(x)cn(x)}
{dn^{2}(x)}+U_{s}(x)]\psi(x)=E\psi(x),
\hspace{0.5cm}U_{s}(x)=K_{2}s(s+1)cd^{2}(x) \eqnum{A7}
\end{equation}
The Hamilton operator can be written as
\begin{equation}
\hat{H}_{s}=K_{1}
[-i\frac{d}{dx}-\tilde{A}(x)]^{2}
-i(s+\frac{1}{2})\lambda\alpha\frac{sn(x)cn(x)}
{dn^{2}(x)}+\tilde {U}_{s}(x) \eqnum{A8}
\end{equation}
The gauge potential
\begin{equation}
\tilde{A}(x)=\frac{\alpha}{2K_{1}dn(x)} \eqnum{A9}
\end{equation}
leads exactly to the
  semiclassical phase angle $\theta_{s}$,
while the scalar potential
 which contains an imaginary part
 is ill-defined. In the large s limit 
one might neglect the imaginary 
part in comparison with the term $U_{s}(x)$ and then has 
 the Hamilton operator given by
\begin{equation}
\hat{H}_{s}=K_{1}[-i\frac{d}{dx}
-\tilde{A}(x)]^{2}
+\tilde{U}_{s}(x), \hspace{0.5cm}
 \tilde{U}_{s}(x)=K_{2}s(s+1)cd^{2}(x)-\frac {\alpha^{2}}{4K_{1}dn^{2}(x)} \eqnum{A10}
\end{equation}
 The final Hamiltonian
\begin{equation}
\tilde{H}_{s}=\frac{1}{2m}
[p-\tilde{A}(x)]^{2}
+\tilde{U}_{s}(x),\hspace{0.5cm}m=\frac{1}{2K_{1}} \eqnum{A11}
\end{equation}
 is the counterpart of the
  effective Hamiltonian $H_{s}$ of eq.(3).
 The corresponding Euclidean Lagrangian is
\begin{equation}
\tilde{L}^{e}_{s}=\frac{m}{2}\dot{x}^{2}+i\tilde{A}(x)\dot{x}+\tilde{U}_{s}(x) \eqnum{A12}
\end{equation}

\end{document}